\documentclass[12pt,a4paper]{article}
\usepackage{color}
\usepackage{cite}
\usepackage{ulem}
\usepackage{latexsym,color}
\usepackage{epsfig}
\usepackage{amssymb,amsmath}
\newcommand{\bc}{\begin{center}}
\newcommand{\ec}{\end{center}}
\newcommand{\bd}{\begin{displaymath}}
\newcommand{\ed}{\end{displaymath}}
\newcommand{\be}{\begin{equation}}
\newcommand{\ee}{\end{equation}}
\newcommand{\ba}{\begin{array}}
\newcommand{\ea}{\end{array}}
\newcommand{\bt}{\begin{tabular}}
\newcommand{\et}{\end{tabular}}

\newcommand{\ds}{\displaystyle}

\begin{document}

\title{Spin-independent interactions of Dirac Fermionic Dark Matter in the composite Higgs models}

\author{M.~G.~Belyakova\,,\quad R.~Nevzorov\\[5mm]
\itshape{I. E. Tamm Department of Theoretical Physics,}\\[0mm]
\itshape{Lebedev Physical Institute, Leninsky prospect 53, 119991 Moscow, Russia}}

\date{}

\maketitle

\begin{abstract}{
\noindent
According to recent measurements, dark matter magnetic dipole moment is strongly constrained.
In the composite Higgs models the magnetic dipole moment of the Dirac dark matter fermion and
its mass can be suppressed by the approximate $U(1)$ symmetry.
We consider $E_6$ inspired composite Higgs model (E$_6$CHM) with $U(1)$ symmetry violating
operators, which give rise to dark matter's mass and coupling constant to Higgs boson.
The dependence of the spin-independent dark matter-nucleon scattering cross section on the
E$_6$CHM parameters is explored. We argue that there are regions of the parameter space which
are still safe from all current constraints and may lead to spectacular LHC signatures.}
\end{abstract}

\newpage
\section{Introduction}

Numerous astrophysical and cosmological observations confirm the presence of a relatively large
non--luminous (dark) component in the matter content of the Universe.
The explanation of these observations requires the existence of galaxy
halos composed of dark matter (DM) objects. Most likely the corresponding non--luminous component
is constituted of nonrelativistic particles which interact gravitationally and possibly by weak interactions.
Candidates for DM state occur naturally in a variety of extensions of the standard model (SM)
and may possess diverse properties. All such states can contribute to DM content of the Universe.

It is especially interesting to explore the nature of DM candidates within well motivated
extensions of the SM that permit to almost stabilize the hierarchy between Grand Unification
(or Planck) and electroweak (EW) scales. Here we focus on the composite Higgs models. They are based on the ideas which
were proposed in the 70's \cite{Terazawa:1976xx} and 80's \cite{composite-higgs}. These extensions of the SM contain two
sectors \cite{Bellazzini:2014yua}. The weakly--coupled sector involves elementary states with the quantum numbers of
the SM fermions and SM gauge bosons. The second, strongly interacting sector lead to a set of resonances. In particular, it is expected
that the breakdown of an approximate global symmetry near the scale $f$ in this sector gives rise to pseudo--Nambu--Goldstone
bosons (pNGBs). The present experimental data indicate that the compositeness scale $f$ should be substantially larger
than $1$ TeV.  Four of the pNGBs form Higgs doublet $H$. The vacuum expectation value (VEV) $v\simeq 246\,\mbox{GeV}$ of $H$
breaks the EW symmetry inducing the masses of all SM particles.

In principle, the strongly interacting sector in the composite Higgs models may also result in the neutral Dirac fermion $\chi$
that can contribute to the DM in our Universe. Such fermion tends to have a magnetic dipole moment $\mu_{\chi}$ that within
these extensions of the SM is of the order of $\mu_{\chi}\sim e/f$ where $e$ is the electron charge. The limits
on $\mu_{\chi}$ set by DM direct detection experiments imply that $f\gtrsim 10^4\,\mbox{TeV}$
\cite{PICO:2022ohk,PANDAX}\footnote{Other bounds are considerably weaker \cite{Hambye:2021xvd}.}.
This bound can be weakened if the Dirac fermion $\chi$ composes only a small fraction $\rho_{\chi}$ of the total
DM density $\rho_{DM}$. Indeed, signal in the direct detection experiments is determined by the concentration (density) of the
DM states and the $\chi$--nuclei scattering cross section. If $\chi$--nuclei scattering is governed by the electromagnetic
interaction then the corresponding cross section is proportional to $\mu_{\chi}^2$. Since the experimental limit on the
magnetic dipole moment $\mu^{exp}_{\chi}$ is set by assuming that fermion $\chi$ constitutes the entirety of the DM,
the bound on $\mu_{\chi}$ in the case $\rho_{\chi}\ll \rho_{DM}$ becomes weaker, i.e.
\be
\mu_{\chi} < \mu^{exp}_{\chi} \left(\dfrac{\rho_{DM}}{\rho_{\chi}}\right)^{1/2}\,.
\label{0}
\ee
Although $\rho_{\chi}$ might be just a tiny fraction of the total DM density $\rho_{DM}$, hereafter we only
consider scenarios with $\left(\dfrac{\rho_{\chi}}{\rho_{DM}}\right) > 10^{-3}$ so that the scale $f$
in the composite Higgs models is still required to be much larger than $100\,\mbox{TeV}$. In this case an enormous
degree of tuning, which is about $\xi\simeq v^2/f^2$, is required to get a Higgs boson with mass $125\,\mbox{GeV}$.
However in some composite Higgs models the interaction in the Lagrangian
\begin{equation}
\mathcal{L}_{MDM}= \frac{\mu_{\chi}}{2} \bar{\chi}_R \,\sigma^{\mu\nu} \chi_L F_{\mu\nu} + h.c.
\label{1}
\end{equation}
can be suppressed by the approximate global symmetry. This symmetry may also ensure that $\chi$ is
the lightest composite state.

In this article we investigate the interaction of the Dirac DM fermion with nucleons in the framework of the $E_6$ inspired composite
Higgs model (E$_6$CHM) \cite{Nevzorov:2015sha}--\cite{Nevzorov:2016fxp}. Within this model the composite sector is invariant under the transformations
of $SU(6)\times U(1)_L\times U(1)_B$ symmetry where $U(1)_L$ and $U(1)_B$ are associated with the conservations of the lepton
and baryon numbers respectively. The breakdown of approximate $SU(6)$ symmetry near the scale $f$ down to $SU(5)$, which involves
the SM $SU(3)_C\times SU(2)_W\times U(1)_Y$ gauge group, gives rise to the composite Higgs doublet $H$. In the limit when the lightest
neutral exotic fermion has a mass which is considerably smaller then $1\,\mbox{TeV}$ the E$_6$CHM possesses an additional approximate
$U(1)_E$ symmetry that allows to suppress the interaction (\ref{1}) in the Lagrangian. We examine the dependence of the
spin-independent part of the DM-nucleon scattering cross section on the parameters of the model in this case and
discuss its possible LHC signatures.

The layout of this paper is as follows. In the next section we briefly review the composite Higgs models
and E$_6$CHM. In section 3 we discuss the generation of masses of exotic fermions.
The dark matter phenomenology is considered in section 4. Section 5 concludes the paper.

\section{Composite Higgs models and E$_6$CHM}

In the minimal composite Higgs model (MCHM) \cite{Agashe:2004rs} the invariance of the Lagrangian of the strongly interacting
sector with respect to $\mbox{SO(5)}\times U(1)_X$ symmetry transformations is imposed. Near the scale $f$ this approximate
global symmetry is broken down to $SO(4)\times U(1)'_X \cong SU(2)_W\times SU(2)_R\times U(1)'_X $, which includes the
$SU(2)_W\times U(1)_Y$ gauge group as a subgroup. The corresponding breakdown leads to four pNGBs that compose
the Higgs doublet $H$. In this case $SU(2)_{cust} \subset SU(2)_W\times SU(2)_R$ \cite{Sikivie:1980hm}
protects the Peskin--Takeuchi $\hat{T}$ parameter \cite{Peskin:1991sw} against the contributions induced by new states.
The contributions of new resonances to the electroweak observables within the composite Higgs models were analysed in Refs. \cite{EWPOCHM}--\cite{Vignaroli:2012si}. The implications of these extensions of the SM were also studied
for Higgs physics \cite{Bellazzini:2012tv}--\cite{Azatov:2013ura}, \cite{Mrazek:2011iu}--\cite{Pomarol:2012qf},
gauge coupling unification \cite{Gherghetta:2004sq}--\cite{Barnard:2014tla}, dark matter \cite{Frigerio:2011zg}, \cite{Frigerio:2012uc},
\cite{Barnard:2014tla}--\cite{Asano:2014wra} and collider phenomenology \cite{Pomarol:2008bh}--\cite{Bellazzini:2012tv},
\cite{Barbieri:2008zt},  \cite{Pomarol:2012qf},  \cite{Redi:2011zi}--\cite{Delaunay:2013pwa}.
Different extensions of the MCHM were considered in Refs. \cite{Frigerio:2011zg}, \cite{Mrazek:2011iu}--\cite{Frigerio:2012uc},
\cite{Barnard:2014tla}--\cite{Asano:2014wra}, \cite{Cacciapaglia:2014uja}.

At low energies the SM fields are superpositions of the corresponding elementary states and their composite partners.
According to this partial compositeness framework \cite{Contino:2006nn, Kaplan:1991dc} the couplings of the SM particles
to the composite Higgs and other composite states are proportional to its fraction of compositeness.
Since all SM fermions except the top quark are relatively light their fractions of compositeness tend to be rather small
resulting in the partial suppression of flavour--changing processes in the composite Higgs models \cite{Contino:2006nn}.
Within these models the constraints caused by the non--diagonal flavour transitions in the quark and lepton sectors were
analysed in Refs. \cite{Barbieri:2012tu}--\cite{Vignaroli:2012si}, \cite{Redi:2011zi},
\cite{Blanke:2008zb}--\cite{Barbieri:2012uh} and \cite{Redi:2013pga}, \cite{Barbieri:2012uh}--\cite{Csaki:2008qq}, respectively.
In general the corresponding constraints require the compositeness scale $f$ to be larger than $10\,\mbox{TeV}$
\cite{Barbieri:2012tu}--\cite{Csaki:2008zd}, \cite{Redi:2011zi}, \cite{Blanke:2008zb}, \cite{Agashe:2006iy}.
Nevertheless in the composite Higgs models with additional global flavour symmetries this bound can be
substantially weakened \cite{Barbieri:2008zt}--\cite{Barbieri:2012tu}, \cite{Redi:2011zi}--\cite{Redi:2013pga},
\cite{Barbieri:2012uh}, \cite{Cacciapaglia:2007fw}.

In general new interactions in the composite Higgs models may give rise to lepton and baryon number violating
processes. Indeed, due to the mixing between the elementary states and their composite partners, the four--fermion operators
resulting in proton decay can be induced through non--perturbative effects. These operators are only suppressed by
the scale $f$ and the small fractions of compositeness of the first and second generation fermions. Such suppression is not
enough to prevent rapid proton decay as well as other baryon number violating processes. Similarly, dimension-5 operators
$\ell_i \ell_j H H/ f$, where $\ell_i$ are $SU(2)_W$ lepton doublets, can be generated leading to far too large
Majorana neutrino masses. Thus in the composite Higgs models one has to impose $U(1)_B$ and $U(1)_L$ symmetries.

The particle content, the global and gauge symmetries of the E$_6$CHM may originate from $E_6\times G_0$
Supersymmetric (SUSY) Grand Unified Theory (GUT). Fields belonging to the weakly--coupled sector participate in the
$E_6$ interactions only whereas strongly interacting sector comprises multiplets which are charged under both the
$G_0$ and $E_6$ gauge groups. Near the GUT scale $M_X\sim 10^{16}\,\mbox{GeV}$ $G_0$ and $E_6$ can be broken to their
subgroups $G$ and $SU(3)_C\times SU(2)_W\times U(1)_Y$ so that $SU(6)$ remains an approximate global symmetry of the
strongly coupled sector \cite{Nevzorov:2015sha,Nevzorov:2022zjo}.

The Lagrangian of the E$_6$CHM may also possess $U(1)_L$ and $U(1)_B$ global symmetries.
Nearly exact conservation of the $U(1)_B$ and $U(1)_L$ charges at low energies implies that the elementary fermions with different
baryon and/or lepton numbers should belong to different $27$--plets of $E_6$. All other components of the corresponding $27$--plets
gain masses of the order of $M_X$. Such splitting of the fundamental representations of $E_6$ can take place within a six--dimensional
orbifold GUT model with $N=1$ SUSY \cite{Nevzorov:2015sha,Nevzorov:2022zjo}. The breakdown of SUSY can occur somewhat below the
GUT scale $M_X$\footnote{The phenomenological aspects of the $E_6$ inspired models with low-scale SUSY breaking were examined in \cite{e6ssm}.}.
In this case the baryon and lepton numbers of the $27$--plets are determined by the $U(1)_B$ and $U(1)_L$ charges ($B$ and $L$) of the
fermion components of these supermultiplets that survive to low energies. Because all components of the $27$--plets carry the same $U(1)_B$
and $U(1)_L$ charges, $E_6$ gauge interactions do not give rise to baryon and lepton number violating operators, in contrast with conventional GUTs.

In principle, near the Planck scale $U(1)_B$ and $U(1)_L$ could be gauge symmetries. These gauge symmetries may be
spontaneously broken so that global $U(1)_B$ and $U(1)_L$ remain almost intact. Within the E$_6$CHM the interactions between the elementary states
and their composite partners break global $SU(6)$ symmetry and its $SU(5)$ subgroup but preserve $SU(3)_C\times SU(2)_W\times U(1)_Y$
gauge symmetry as well as $U(1)_B$ and $U(1)_L$. In other words, at low energies the Lagrangian of the strongly coupled sector of the E$_6$CHM
should be invariant under the transformations of an $SU(6)\times U(1)_B\times U(1)_L$ global symmetry, whereas the full effective Lagrangian
of the E$_6$CHM respects $SU(3)_C\times SU(2)_W\times U(1)_Y \times U(1)_B\times U(1)_L$ symmetry.

The $U(1)_B$ and $U(1)_L$ symmetries permit to get the appropriate suppression of the proton decay rate
in the E$_6$CHM. Hereafter we assume that $U(1)_B$ is almost exact. The approximate $U(1)_L$ symmetry forbids operators
that give rise to unacceptably large masses of the left--handed neutrino in the composite Higgs models.
Tiny Majorana masses of the left--handed neutrino can be still induced if in the weakly--coupled sector
$U(1)_L$ is broken down to $Z^L_{2}=(-1)^{L}$, where $L$ is a lepton number.

Below the scale $f$ the approximate $SU(6)$ symmetry is broken down to $SU(5)$ giving rise to eleven pNGB states.
The $SU(5)$ symmetry does not contain $SU(2)_{cust}$ subgroup. As a consequence the electroweak precision measurements
set stringent lower limit $f\gtrsim 5-6\,\mbox{TeV}$ in the E$_6$CHM \cite{Nevzorov:2015sha}.
The generators of the $SU(5)$ subgroup of $SU(6)$ and eleven generators from the coset $SU(6)/SU(5)$ are denoted here
by $T^a$ and $T^{\hat{a}}$ respectively, where $\mbox{Tr}\Biggl(T^a T^b\Biggr) = \ds\frac{1}{2} \delta_{ab}$,
$\mbox{Tr} \Biggl(T^{\hat{a}} T^{\hat{b}}\Biggr) = \ds\frac{1}{2} \delta_{\hat{a}\hat{b}}$ and $\mbox{Tr}\Biggl(T^a T^{\hat{b}}\Biggr)= 0$.
For the eleven pNGB states it is convenient to use the following non--linear representation \cite{Nevzorov:2015sha}
\be
\ba{c}
\Omega^T = \Omega_0^T \Sigma^T = e^{i\frac{\phi_0}{\sqrt{15}f}}
\Biggl(C \phi_1\quad C \phi_2\quad C \phi_3\quad C \phi_4\quad C\phi_5\quad \cos\dfrac{\tilde{\phi}}{\sqrt{2} f} + \sqrt{\dfrac{3}{10}} C \phi_0 \Biggr)\,,\\[3mm]
C=\dfrac{i}{\tilde{\phi}} \sin \dfrac{\tilde{\phi}}{\sqrt{2} f}\,,\qquad \tilde{\phi}=\sqrt{\dfrac{3}{10}\phi_0^2+|\phi_1|^2+|\phi_2|^2+|\phi_3|^2+|\phi_4|^2+|\phi_5|^2}\,,
\ea
\label{2}
\ee
where a 6--component unit vector $\Omega$ is a fundamental representation of $SU(6)$ given by
$$
\Omega^T = \Omega_0^T \Sigma^T\,,\qquad
\Omega_0^T= (0\quad 0\quad 0\quad 0\quad 0\quad 1)\,,\qquad \Sigma= e^{i\Pi/f}\,,\qquad \Pi=\Pi^{\hat{a}} T^{\hat{a}}\,.
$$
In Eq.~(\ref{2}) fields $\phi_1\,, \phi_2\,, \phi_3\,, \phi_4$ and $\phi_5$ are complex whereas $\phi_0$ is a real field.
Because $\phi_0$ and $\tilde{\phi}$ are invariant under the preserved $SU(5)$ vector $\Omega$ transforms
as $\bf{5}+\bf{1}$ under the transformation of the unbroken $SU(5)$, where a 5--component vector is formed by
$\bf{5}\sim (\phi_1\,\, \phi_2\,\, \phi_3\,\, \phi_4\,\, \phi_5)$. The $SU(5)$ singlet state $\bf{1}=\phi_0$ is a real SM singlet field.
The first two components of $\bf{5}$, i.e. $(\phi_1\, \phi_2)$, transform as an $SU(2)_W$ doublet and therefore they are associated
with the SM--like Higgs doublet $H$. Three other components of $\bf{5}$, i.e. $(\phi_3\, \phi_4\, \phi_5)$, form an $SU(3)_C$ triplet $T$.
Since Higgs doublet has $B=L=0$, all components of $\Omega$ do not carry any lepton and/or baryon numbers.

In the leading approximation the Lagrangian describing the interactions of the pNGBs can be written as
\be
\mathcal{L}_{pNGB}=\ds\dfrac{f^2}{2}\biggl|\mathcal{D}_{\mu} \Omega \biggr|^2\,.
\label{3}
\ee
The effective potential $V(H, T, \phi_0)$, that vanishes in the exact $SU(6)$ symmetry limit,
is induced by the interactions of the elementary fermions and gauge bosons with their composite partners,
which break global $SU(6)$ symmetry. The analysis performed within the composite Higgs models, which are similar to the E$_6$CHM,
indicates that there is a substantial region of the parameter space where the EW symmetry is broken to $U(1)_{em}$, corresponding
to electromagnetism, while the $SU(3)_C$ symmetry is preserved \cite{Frigerio:2011zg},\cite{Barnard:2014tla}. It was also shown
that in such models the parameters may be tuned so that $125\,\mbox{GeV}$ Higgs state can be obtained \cite{Barnard:2014tla}.
The $SU(3)_C$ triplet scalar $T$ is much heavier than the SM--like Higgs scalar in this case.

\section{Exotic fermions in the E$_6$CHM}

Since $t$--quark is considerably heavier than other SM fermions, the compositeness fraction of the right--handed top
quark $t^c$ should be of the order of unity. In fact, the E$_6$CHM implies that $t^c$ is entirely composite. This can happen if
the weakly--coupled sector includes the following set of fermion multiplets\cite{Nevzorov:2015sha}
\begin{equation}
(q_i,\,d^c_i,\,\ell_i,\,e^c_i) + u^c_{\alpha} + \bar{q}+\bar{d^c}+\bar{\ell}+\bar{e^c} \, ,
\label{4}
\end{equation}
where $\alpha=1,2$ and $i=1,2,3$. In Eq.~(\ref{4}) $e_i^c$, $u_{\alpha}^c, d_i^c$ represent the right-handed charged leptons,
up- and down-type quarks, $\ell_i$ and $q_i$ are associated with the left-handed lepton and quark doublets, whereas
extra exotic fermions $\bar{e^c}$, $\bar{\ell}$, $\bar{q}$ and $\bar{d^c}$ have exactly opposite $SU(3)_C\times SU(2)_W\times U(1)_Y$
quantum numbers to the right-handed charged leptons, left-handed lepton doublets, left-handed quark doublets and right-handed down-type quarks,
respectively. The set of multiplets (\ref{4}) is chosen so that the weakly--coupled sector contains all SM fermions except $t^c$ and
anomaly cancellation takes place. It is expected that the dynamics of strongly interacting sector results in the ${\bf 10} + {\bf \overline{5}}$
fermion multiplets of $SU(5)$ below the compositeness scale $f$. Most components of these $SU(5)$ multiplets get combined with $\bar{e^c}$,
$\bar{\ell}$, $\bar{q}$ and $\bar{d^c}$ leading to vector--like fermions with masses of the order of $f$. However the components of the $10$--plet
associated with the right--handed top quark $t^c$ survive to the EW scale. The particle content of the weakly--coupled sector (\ref{4})
leads to the approximate unification of the SM gauge couplings near the scale $M_X\sim 10^{15}-10^{16}\, \mbox{GeV}$.

The composite ${\bf 10} + {\bf \overline{5}}$ multiplets of $SU(5)$ can stem from one
${\bf{15}}$--plet and two ${\bf \overline{6}}$--plets (${\bf \overline{6}}_1$ and ${\bf \overline{6}}_2$) of $SU(6)$
that have the following decomposition in terms of $SU(5)$ representations:
${\bf 15}={\bf 10} \oplus {\bf 5}$ and ${\bf \overline{6}}={\bf \overline{5}} \oplus {\bf 1}$.
The components of ${\bf{15}}$, ${\bf \overline{6}}_1$ and ${\bf \overline{6}}_2$ decompose
under $SU(3)_C\times SU(2)_W\times U(1)_Y\times U(1)_B$ as follows:
\begin{equation}
\ba{ll}
\ba{rcl}
{\bf 15} &\to& Q = \left(3,\,2,\,\dfrac{1}{6},\,-\dfrac{1}{3}\right)\,,\\[2mm]
&& t^c = \left(\bar{3},\,1,\,-\dfrac{2}{3},\,-\dfrac{1}{3}\right)\,,\\[2mm]
&& E^c = \Biggl(1,\,1,\,1,\,-\dfrac{1}{3},\, -\dfrac{1}{3}\Biggr)\,,\\[2mm]
&& D = \left(3,\,1,\,-\dfrac{1}{3},\,-\dfrac{1}{3} \right)\,,\\[2mm]
&& \overline{L}=\left(1,\,2,\,\dfrac{1}{2},-\dfrac{1}{3}\,\right)\,;
\ea
\qquad
\noindent
\ba{rcl}
{\bf \overline{6}}_{1} &\to & D^c_{1} = \left(\bar{3},\,1,\,\dfrac{1}{3},\,\dfrac{1}{3} \right)\,,\\[2mm]
& & L_{1} = \left(1,\,2,\,-\dfrac{1}{2},\,\dfrac{1}{3} \right)\,,\\[2mm]
& & N_{1} = \Biggl(1,\,1,\,0,\,\dfrac{1}{3} \Biggr)\,;\\[6mm]
{\bf \overline{6}}_{2} &\to & D^c_{2} = \left(\bar{3},\,1,\,\dfrac{1}{3},\,-\dfrac{1}{3} \right)\,,\\[2mm]
& & L_{2} = \left(1,\,2,\,-\dfrac{1}{2},\,-\dfrac{1}{3} \right)\,,\\[2mm]
& & \overline{N}_{2} = \Biggl(1,\,1,\,0,\,-\dfrac{1}{3} \Biggr)\,.
\ea
\ea
\label{5}
\end{equation}
The first and second quantities in brackets are the $SU(3)_C$ and $SU(2)_W$ representations, whereas the third
and fourth quantities are $U(1)_Y$ and $U(1)_{B}$ charges.

The right--handed top quark $t^c$ is contained in ${\bf{15}}$--plet of $SU(6)$. Then the baryon number conservation
requires all components of this multiplet to carry the same baryon number $B=-1/3$.
The composite partners $Q_3$ of the left--handed top quark and left--handed $b$--quark, i.e. $q_3$, belong to ${\bf 20}(Q_3)$
representations of $SU(6)$, that has the following decomposition in terms of $SU(5)$ representations:
${\bf 20}={\bf 10} \oplus {\bf \overline{10}}$. The simplest $SU(6)$ generalisation of the Yukawa interactions,
that can give rise to the top quark mass, can be written in the following form \cite{Nevzorov:2015sha}:
$$
\mathcal{L}^t_{SU(6)}\sim {\bf 20}(Q_3)\times {\bf 15}\times \Omega + h.c.\,.
$$

The masses of $D$ and $\overline{L}$ components of ${\bf{15}}$--plet are induced through interaction
\be
\mathcal{L}^d_{SU(6)}= h_{N} f ({\bf 15}\times {\bf \overline{6}_1}\times \Omega^{\dagger}) + h.c.\,.
\label{6}
\ee
The dimensionless coupling $h_{N}$ is expected to be of the order of unity.
The corresponding operator is allowed only if all components of $\overline{6}_1$ have baryon number $B=1/3$.
The non--zero mass of $N_{1}$ and $\overline{N}_{2}$ can be induced if the baryon number of $\overline{6}_2$ is either
$1/3$ or $(-1/3)$. Here we assume that $D^c_{2}$, $L_{2}$ and $N_{2}$ carry baryon number $B=-1/3$. Then
the corresponding mass term is generated after the breakdown of $SU(6)$ symmetry through interaction
\be
\mathcal{L}^n_{SU(6)}= g_{N} f ({\bf \overline{6}_1}\times \Omega)({\bf \overline{6}_2}\times \Omega) + h.c.\,,
\label{7}
\ee
where $g_{N}$ is a dimensionless coupling. As a result the full set of mass terms associated with
exotic states in the E$_6$CHM can be written as follows
\be
\mathcal{L}_{mass} = \mu_{q} \bar{q} Q + \mu_e \bar{e^c} E^c + \mu_{D} D^c_{1} D + \mu_{L} \overline{L} L_{1}
+ \mu_d \bar{d^c} D^c_{2} + \mu_{l} \bar{\ell} L_{2} + \mu_N \overline{N}_{2} N_{1} + h.c.\,,
\label{8}
\ee
where $\mu_{N} \simeq g_{N} f$, $\mu_{D}\simeq \mu_{L}\simeq h_{N} f$ and $\mu_{q}\sim \mu_e \sim \mu_d \sim \mu_{l} \sim f$.
All exotic fermions mentioned above do not carry any lepton number.

Eq.~(\ref{8}) indicates that in the limit $g_{N}=0$ the E$_6$CHM Lagrangian is invariant under the transformations of
$U(1)_E$ global symmetry defined as
\be
{\bf \overline{6}_2} \longrightarrow e^{i\beta} {\bf \overline{6}_2},\qquad \bar{d^c} \longrightarrow e^{-i\beta} \bar{d^c},\qquad
\bar{\ell} \longrightarrow e^{-i\beta} \bar{\ell}\,.
\label{9}
\ee
The approximate $U(1)_E$ symmetry ensures that the lightest exotic fermion $\chi$ is mostly superposition of $N_{1}$ and $N_{2}$.
The baryon number conservation does not permit such lightest exotic state to decay. This can be understood in terms of the
discrete $Z_3$ symmetry which is known as baryon triality \cite{Frigerio:2011zg}, \cite{baryon-triality}.
The corresponding symmetry transformations are given by
\be
\Psi \longrightarrow e^{2\pi i B_3/3} \Psi,\qquad B_3 = (3 B - n_C)_{\mbox{mod}\,\, 3}\,,
\label{10}
\ee
where $B$ is the baryon number of the multiplet $\Psi$ and $n_C$ is the number of colour indices ($n_C=1$ for the
colour triplet and $n_C=-1$ for antitriplet). Since the baryon number is preserved, the E$_6$CHM Lagrangian is also invariant
under the symmetry transformations (\ref{10}). All SM bosons and SM fermions have $B_3=0$ while exotic fermions
and the scalar colour triplet $T$ carry either $B_3=2$ or $B_3=1$. As a consequence the lightest exotic fermion $\chi$
with non--zero $B_3$ charge cannot decay into SM particles and should therefore be stable.

\section{Dark matter phenomenology}

\subsection{Dark matter searches}

Thus in the E$_6$CHM $\chi$ may account for all or some of the observed cold DM density.
Since in this model the compositeness scale $f$ is larger than $5\,\mbox{TeV}$ we restrict our
consideration here to the scenarios in which the mass $m_{\chi}$ of the lightest exotic fermion $\chi$
exceeds $200\,\mbox{GeV}$. Nowadays DM particles are currently actively searched for at colliders
and in dedicated direct and indirect detection experiments.

Observation of the rotation curves (RC) gives an opportunity to investigate mass distribution of
luminous and dark components of the galaxy. The sophisticated simulations of dark matter density
profile have been conducted to find the one which resembles RC observations. The last claims that outer
region of galaxies are dark--matter--dominated. Analysis of rotational velocities indicates the
presence of DM core in the center of the galaxies. The intrinsic DM distribution has not been defined yet.
Independent of any particular DM distribution profile
the dark matter density in the Solar system is about $0.4\,\mbox{GeV}/\mbox{cm}^3$ \cite{Catena:2009mf}.
Galactic halos are considered to be fixed in comparison to the rotating disc of the galaxy while Sun and Earth
move in the rest frame of the galaxy with a speed of about $200\,\mbox{km}/\mbox{s}$. If DM were composed
of particles with masses around $100\,\mbox{GeV}$ then, seen from Earth, they would have a relative velocity
of about $200\,\mbox{km}/\mbox{s}$ and during one year each cubic meter on Earth would be crossed by $\sim 10^{13}$
DM particles \cite{dm-review}. Building large detectors one may hope to explore the DM
interactions with ordinary matter via scattering with the nucleons (or electrons) of the atoms present
in the experiments. In the direct detection experiments the recoil energy of nuclei is measured in order
to estimate the dark matter scattering cross section with nucleons that involves the spin--dependent and
spin-independent parts.

Several direct detection experiments, including LUX--ZEPLIN experiment (LZ), the XENON dark matter
research project (XENONnT) as well as the Particle and Astrophysical Xenon Detector (PandaX-4T),
are currently running. Assuming that DM is formed by a single type of particles
these experiments set most stringent constraints on the spin--independent DM--nucleon
scattering cross section $\sigma_{\text{exp}}$ \cite{PandaX-4T:2021bab,XENON:2023cxc,LZ}.
The corresponding experimental limits can be substantially relaxed
in the E$_6$CHM if the lightest exotic fermion $\chi$ gives only a minor contribution $\rho_{\chi}$ to the total
DM density $\rho_{DM}$. In this case the computed spin--independent DM--nucleon
scattering cross section $\sigma$ should satisfy the restriction
\be
\sigma < \sigma_{\text{exp}}\left(\dfrac{\rho_{DM}}{\rho_{\chi}}\right)\,.
\label{101}
\ee
In the next subsection we compare the results of our calculations of $\sigma$ with the present
experimental bounds.

Indirect detection of dark matter refers to the search for the annihilation products from
DM particles which can be detected on Earth. The main types of such annihilation products are
gamma rays (photons), neutrinos, and charged particles (protons, antiprotons, electrons and positrons).
These particles tend to come from the regions where dark matter is denser.

Photons are the easiest to detect. Their propagation is almost unaffected by the intergalactic and
interstellar media. DM annihilation to almost any pair of SM states give rise to gamma rays.
In particular, given the possibility of annihilation of the DM particles into two photons,
the cosmic photon spectrum should contain a line at energies equal to the mass of these states.
Although most intense gamma-ray emission from the DM annihilation must take place near the
Galactic center, because it hosts the highest density of dark matter, this region is also very bright
making the extraction of a signal highly problematic. Nearby dwarf spheroidal galaxies have rather
large dynamical mass and contain relatively small amounts of gas so that they are an optimal target
choice for DM searches. The explorations of the gamma--ray sky with the space observatory
Fermi Gamma--Ray Space Telescope \cite{Fermi-LAT:2015att, Fermi-LAT:2016uux, MAGIC:2016xys}
and with ground--based facilities such as Major Atmospheric Gamma Imaging Cherenkov
Telescopes (MAGIC) \cite{MAGIC:2016xys, MAGIC:2022acl}, High Energy Stereoscopic
System (H.E.S.S.) \cite{HESS:2016mib, HESS:2022ygk}, Very Energetic Radiation Imaging
Telescope Array System (VERITAS) \cite{VERITAS:2017tif, VERITAS:2024usn} and
High Altitude Water Cherenkov Observatory (HAWC) \cite{HAWC:2017mfa, HAWC:2023owv}
allowed to set limits on the annihilation cross sections of DM
particles with masses from a few GeV up to $10\,\mbox{TeV}$. Somewhat weaker limits
on this cross section were set by the IceCube Neutrino Observatory (IceCube)
\cite{IceCube:2017rdn, IceCube:2023ies, ANTARES:2020leh} as well as
Astronomy with a Neutrino Telescope and Abyss environmental Research project (ANTARES)
\cite{ANTARES:2020leh, ANTARES:2015vis} since searches for an anomalous flux of
high--energy neutrinos yielded null results.

Charged particles are strongly affected by the magnetic fields and interstellar medium
so that their propagation is quite complex to model. Nevertheless stringent constraint
on the DM annihilation cross section was obtained in \cite{Cuoco:2017iax}
using the antiproton data measured by Alpha Magnetic Spectrometer (AMS-02) \cite{AMS:2016oqu}.
Energy injection from the DM annihilation in the early Universe affects the
anisotropies of the cosmic microwave background (CMB). The observations made by the
space observatory Planck also permit to set upper bounds on the corresponding
cross section \cite{Planck:2018vyg,Kawasaki:2021etm}. It is expected that Cherenkov Telescope Array (CTA)
is going to widen the reach of DM searches in the foreseeable future \cite{CTA:2020qlo, Hiroshima:2019wvj}.

In the E$_6$CHM the annihilation of DM fermions is strongly suppressed due to the small or negligible
number of DM antiparticles left over from the early Universe. This allows to avoid the
existing indirect detection limits and the constraints on the annihilation of DM states into
electromagnetically charged particles during the CMB era. Indeed, in the E$_6$CHM the lightest exotic
fermion carries baryon number. Therefore the relic abundance of the DM states
in the model under consideration should be induced by the same mechanism that results
in the baryon asymmetry\footnote{Within the present version of the E$_6$CHM such mechanism has not been proposed yet.
Nevertheless the baryon asymmetry generation was considered in other modification of this model
\cite{Nevzorov:2022zjo,Nevzorov:2017rtf}.}. Thus the lightest exotic fermions form asymmetric dark matter \cite{Petraki:2013wwa}.
In the visible sector of the SM the strong baryon--antibaryon annihilation drives the abundance of antibaryons
to almost zero, leaving only the excess of baryons present today. Similarly, the "symmetric part of DM states"
in the E$_6$CHM, i.e. the lightest exotic antifermions plus an equal number of the lightest exotic fermions, might be
efficiently annihilated away in the early Universe if there exist sufficiently strong interactions that permit
such annihilation. This type of process requires a DM annihilation cross section which is just a few times larger
than the canonical value for the thermally averaged cross section that yields the observed DM abundance of
weakly--interacting massive particles  \cite{Petraki:2013wwa}, i.e.
$\langle \sigma v\rangle \simeq 2.3\times 10^{-26} \mbox{cm}^3/\mbox{s}$ \cite{Steigman:2012nb}.
Within the E$_6$CHM so efficient $\chi\overline{\chi}$ annihilation into SM particles
can happen if the mass of the lightest exotic fermion $m_{\chi}$ is quite
close to half the mass of the SM singlet pNGB state $\phi_0$, i.e. $m_A/2$. In this scenario some part of
the baryon asymmetry $B_{\chi}$ of our Universe is stored in the DM sector of the E$_6$CHM.
The ratio of the baryon charges $B_{\chi}$ and $B_N$ accumulated by $\chi$ and nucleons can be estimated as
\be
\dfrac{B_{\chi}}{B_N}\simeq \dfrac{1}{3} \left(\dfrac{\rho_{\chi}}{\rho_N}\right) \left(\dfrac{m_N}{m_{\chi}}\right)\,,
\label{11}
\ee
where $\rho_N$ and $\rho_{\chi}$ are contributions of nucleons and lightest exotic fermions to the total energy density,
while $m_N$ is a mass of nucleons.


If dark matter interacts with nucleons, it can scatter off the nuclei in the stars, lose energy and get captured in
interiors of the stars. When dark matter is asymmetric, DM particle--antiparticle annihilation is suppressed
and the DM density in the stars keeps increasing over time. In particular, this type of DM is captured
efficiently in compact objects, such as white dwarfs and neutron stars. The denser accumulation of asymmetric dark matter
may change the thermal evolution of stars or even eventually reach the critical density in the central region of the stars
resulting in their gravitational collapse into black holes. The observation of old neutron stars in DM--rich environments
can thus constrain the mass and interaction cross section of asymmetric dark matter with nucleons \cite{Goldman:1989nd, Kouvaris:2010jy}. In the next
subsection we explore the dependence of the couplings of the lightest exotic fermions to the SM Higgs scalar and $Z$--boson.
These couplings are quite small so that the constraints, that come from the existence of white dwarfs and neutron stars, are satisfied \cite{Kouvaris:2010jy}.

The attractive self--interaction of fermionic asymmetric dark matter may affect the gravitational collapse
of old neutron stars \cite{Kouvaris:2011gb} (see also \cite{dm-review}). In the E$_6$CHM the self--interaction
of the lightest exotic fermions tends to be too weak to facilitate the collapse of such stars.


At colliders DM particles escape the detection giving rise to significant amounts of missing energy and momentum.
Then the only way to reveal that the DM states have been produced is to observe the accompanying SM particles.
ATLAS and CMS collaborations carried out monojet (monophoton) search, where missing energy (associated with dark matter)
is accompanied by a high energy jet (hard photon) which is produced as initial state radiation \cite{ATLAS:2017bfj}.
Within the E$_6$CHM the lightest exotic fermions are mostly produced at the LHC via the Higgs scalar, photon and $Z$--boson
exchange. Our analysis performed in the next subsection indicates that the corresponding couplings of the lightest
exotic fermions are rather small. Therefore monojet (monophoton) production at the LHC is extremely suppressed.

Nevertheless E$_6$CHM implies that the scalar colour triplet $T$ may be lighter than $2\,\mbox{TeV}$ when $f\simeq 5-6\,\mbox{TeV}$.
If such states do exist at so low scales, they can be accessed at the LHC. In collider experiments these scalars
can only be created in pairs. At the LHC the pairs of $T\overline{T}$ are mostly produced through the gluon fusion.
After being produced $T$ sequentially decays into $b$--quark and $\overline{\chi}$ giving rise to some enhancement
of the cross section of
\begin{equation}
pp\to b\overline{b} + E^{\rm miss}_{T} + X\,,
\label{24}
\end{equation}
where $E^{\rm miss}_{T}$ is the missing energy and transverse momentum associated with $\chi$ and $\overline{\chi}$ in the final state
while $X$ can include extra charged leptons and/or jets that may come from the decays of intermediate states.
Similar enhancement of the cross section of the process (\ref{24}) can be caused by the presence of
relatively light superpartner of the $b$--quark. If the mass of the DM fermion is smaller than $800\,\mbox{GeV}$
then ATLAS experiment excluded the bottom squarks with masses below $1200-1270\,\mbox{GeV}$ at $95\%$ confidence level \cite{ATLAS:2021yij}.
Limits from CMS are comparable \cite{CMS:2019ybf}. At the same time no limit can be placed for the DM fermion mass above $800\,\mbox{GeV}$.
The experimental bound mentioned here are fully applicable to the case of the scalar colour triplet $T$.

\subsection{Spin-independent interactions of dark matter with nucleons in the E$_6$CHM}

In the composite Higgs models with $f\gg v$ the spin--independent interactions of the Dirac fermionic Dark Matter with nucleons
are mostly mediated by the $t$-channel exchange of the $125\,\mbox{GeV}$ Higgs scalar, $Z$--boson and photon.
Here, motivated by the nonobservation of CP violation beyond the SM, invariance under CP transformation is imposed.
In this case the $SU(5)$ singlet pNGB state $\phi_0$ manifests itself in the Yukawa interactions with fermions as a pseudoscalar field.
As a consequence $\phi_0$ can not mix with the Higgs boson and does not contribute to the spin--independent interactions of $\chi$
with nucleons. Other composite resonances in the E$_6$CHM have masses which are considerably larger than $10\,\mbox{TeV}$, so that
they are too heavy to be probed at the LHC. Besides their couplings to almost all quarks and leptons are very suppressed since
the fractions of compositeness of almost all SM fermions are rather small. Therefore the contributions of heavy composite resonances
to the spin--independent interactions of $\chi$ with nucleons can be neglected in the leading approximation.

Moreover the contributions of diagrams involving virtual photon or virtual Higgs state may be also strongly suppressed by
the approximate global $U(1)_E$ symmetry. To simplify our analysis here we assume that the magnetic dipole moment of the lightest
exotic fermion $\mu_{\chi}$ is much smaller than the present experimental limit $\sim 10^{-8}\,\mbox{GeV}^{-1}$ and
the interaction (\ref{1}) can be ignored. At the same time we allow the $U(1)_E$ symmetry violating operators that give rise
to $m_{\chi}$ and the coupling of $\chi$ to the $125\,\mbox{GeV}$ Higgs boson.

The $SU(2)_W$ doublets $\bar{\ell}$, $\overline{L}$, $L_{1}$ and $L_{2}$ contain electrically neutral components
$\bar{\nu}_{\ell}$, $\overline{\nu}_L$, $\nu_{1}$ and $\nu_{2}$. These components mix with $N_{1}$ and $\overline{N}_{2}$
after the breakdown of the EW symmetry. When $U(1)_E$ symmetry is preserved to a very good approximation
the conservation of the baryon number allows only a few terms in the E$_6$CHM Lagrangian that give rise to such mixing, i.e.
\be
\mathcal{L}_{mix} = h_N (\overline{L} H^{c}) N_1 + \tilde{h}_N (\bar{\ell} H^{c}) \overline{N}_{2} + h.c.\,.
\label{12}
\ee
In this limit the left-handed and right-handed components of the lightest exotic fermion $\chi$ are given by
\be
\chi_L \simeq N_1 \cos\theta_1  - \nu_1 \sin\theta_1\,,\qquad \chi_R \simeq N_2 \cos\theta_2 - \bar{\nu}_{2} \sin\theta_2\,,
\label{13}
\ee
where $\tan\theta_1\simeq \dfrac{v}{\sqrt{2} f}$ and $\tan\theta_2 \simeq \dfrac{\tilde{h}_N v}{\sqrt{2} \mu_{\ell}}$.
The mixing angles $\theta_1$ and $\theta_2$ determine the strength of the interaction of $\chi$ with the $Z$--boson.
The corresponding part of the Lagrangian can be written as
\be
\ba{c}
\mathcal{L}_{Z\chi}=\overline{\chi} (a^{\chi}_{V} \gamma^{\mu} + a^{\chi}_{PV} \gamma^{\mu} \gamma^{5})\chi Z_{\mu}\,,\\[1mm]
a^{\chi}_{V}=\dfrac{\bar{g}}{4}(\sin^2 \theta_1 - \sin^2 \theta_2)\,,\qquad\qquad
a^{\chi}_{PV}=\dfrac{\bar{g}}{4}(\sin^2 \theta_1 + \sin^2 \theta_2)\,.
\ea
\label{14}
\ee
In Eq.~(\ref{14}) $\bar{g}=\sqrt{g^2+g^{'2}}$ whereas $g$ and $g'$ are $SU(2)_W$ and $U(1)_Y$ gauge couplings.
Since $v\ll f$ the mixing angles $\theta_1$ and $\theta_2$ as well as the couplings $a^{\chi}_{V}$ and $a^{\chi}_{PV}$
are always small. Therefore it is convenient to use the following parameterisation
\be
\ba{ll}
a^{\chi}_{V} = \dfrac{\bar{g} v^2}{8 f^2} c^{\chi}_{V}\,,\qquad\qquad & c^{\chi}_{V}\simeq 1 - \Biggl(\dfrac{\tilde{h}_N f}{\mu_{\ell}}\Biggr)^2\,,\\[1mm]
a^{\chi}_{PV}= \dfrac{\bar{g} v^2}{8 f^2} c^{\chi}_{PV}\,,\qquad\qquad & c^{\chi}_{PV}\simeq 1 + \Biggl(\dfrac{\tilde{h}_N f}{\mu_{\ell}}\Biggr)^2\,.
\ea
\label{15}
\ee
When $\mu_{\ell}\gtrsim \tilde{h}_N f$, the dimensionless coefficients $c^{\chi}_{V}$ and $c^{\chi}_{PV}$ are of the order of unity.
The interactions (\ref{14})--(\ref{15}) may also originate from the non--renormalizable terms in the E$_6$CHM Lagrangian
\be
\mathcal{L}_{Z\chi}=\dfrac{1}{f^2}(H^{\dagger} i D_{\mu} H) \Biggl(a_1 \overline{N}_{1} \gamma^{\mu} N_{1}
+ a_2 \overline{N}_{2} \gamma^{\mu} N_{2} \Biggr)\,,
\label{102}
\ee
where $D_{\mu}$ is a covariant derivative, while absolute values of the parameters $a_1$ and $a_2$ are expected to be of
the order of unity.

Using the Lagrangian describing the interactions of the $Z$--boson with quarks
\be
\ba{c}
\mathcal{L}_{Zq} = \sum_q \dfrac{\bar{g}}{2}\, \overline{q} (a^{q}_{V} \gamma^{\mu} + a^{q}_{PV} \gamma^{\mu} \gamma^{5})q \, Z_{\mu}
=\dfrac{\bar{g}}{2} J^{\mu}_{NC} Z_{\mu}\,,\\[1mm]
a^{q}_{V}=T_{3q} - 2 s^2_W Q_q\,,\qquad\qquad a^{q}_{PV}=T_{3q}\,,
\ea
\label{16}
\ee
one can compute in the leading approximation the hadronic matrix elements
\be
\ba{c}
\langle N' | J^{\mu}_{NC} |N \rangle =  \overline{\Psi}'_N (a^{N}_{V} \gamma^{\mu} + a^{N}_{PV} \gamma^{\mu} \gamma^{5}) \Psi_N\,,\\
a^{p}_{V} \simeq \dfrac{1}{2} - 2 s^2_W\,,\qquad a^{p}_{PV} \simeq \sum_{q=u,d,s} T_{3q} \Delta^{p}_q\,,\\
a^{n}_{V} \simeq - \dfrac{1}{2}\,,\qquad\qquad a^{n}_{PV} \simeq \sum_{q=u,d,s} T_{3q} \Delta^{n}_q\,.
\ea
\label{17}
\ee
In Eq.~(\ref{16})--(\ref{17}) $N=p,n$ and $s_W\simeq g'/\bar{g}$ while
$T_{3q}$ and $Q_q$ are the third component of isospin and the electric charge of the quark $q$ respectively.
For fractions of the nucleon spin carried by a given quark $q$ one can use (see also \cite{Chalons:2012xf})
\begin{equation}
\Delta^{p}_u=\Delta^{n}_d=0.842\,,\qquad \Delta^{p}_d=\Delta^{n}_u=-0.427\,,\qquad \Delta^{p}_s=\Delta^{n}_s=-0.085\,.
\label{171}
\end{equation}

In the process of calculation of the couplings (\ref{14})--(\ref{15}) and (\ref{16})--(\ref{17})
we used SM expressions for the couplings of the $Z$--boson to quarks and leptons. On the other hand in the composite
Higgs models one may expect that the corresponding expressions should get modified. For instance, such modifications
occur due to the mixing between the elementary gauge bosons and their composite partners. In this context it is
worth to point out that the deviations of the couplings of the SM gauge bosons are determined by the parameter $v^2/f^2$.
Because in the E$_6$CHM $f\gtrsim 5\,\mbox{TeV}$ these deviations are rather small and can be ignored in the
leading approximation. It seems rather problematic to test so small deviations of couplings at the LHC and even at future
$e^{+}e^{-}$ collider.

The interaction of the $125\,\mbox{GeV}$ Higgs boson $h$ with the lightest exotic fermion is determined
by the following term in the Lagrangian
\be
\mathcal{L}_{H\chi}=\dfrac{\varepsilon_H}{f} H^{\dagger} H (\overline{N}_{2} N_{1}) + h.c.\,.
\label{18}
\ee
The $U(1)_E$ symmetry forbids operator (\ref{18}) so that in this limit the dimensionless parameter $\varepsilon_H$ vanishes.
In the case of approximate $U(1)_E$ symmetry the lightest exotic fermion $\chi$ gains non--zero mass $m_{\chi}\ll f$ while
$\varepsilon_H$ is expected to be small, i.e. $\varepsilon_H \ll 1$. As a consequence the breakdown of the EW symmetry leads to
the small coupling of the $125\,\mbox{GeV}$ Higgs state $h$ to $\chi$
\be
\mathcal{L}_{h\chi\chi}= g_{h\chi\chi} (\overline{\chi} \chi) h\,,\qquad\qquad  g_{h\chi\chi}\simeq \varepsilon_H \dfrac{v}{f}\,.
\label{19}
\ee

The strength of the spin--independent interactions of $\chi$ with nucleons in the E$_6$CHM is defined by two couplings $a^{\chi}_{V}$
and $|g_{h\chi\chi}|$. Instead of these couplings it is more convenient to use $c^{\chi}_{V}$ and $\varepsilon_H$.
Fig.~1 shows the dependence of $a^{\chi}_{V}$ and $|g_{h\chi\chi}|$ on the compositeness scale $f$
for different values of $c^{\chi}_{V}$ and $\varepsilon_H$. From this figure it follows that in the E$_6$CHM
$a^{\chi}_{V}\sim 10^{-4}$ while $|g_{h\chi\chi}|$ tends to be considerably smaller than $0.01$.

\begin{figure}
\begin{center}
	
\hspace*{-2em}\includegraphics[scale=0.85]{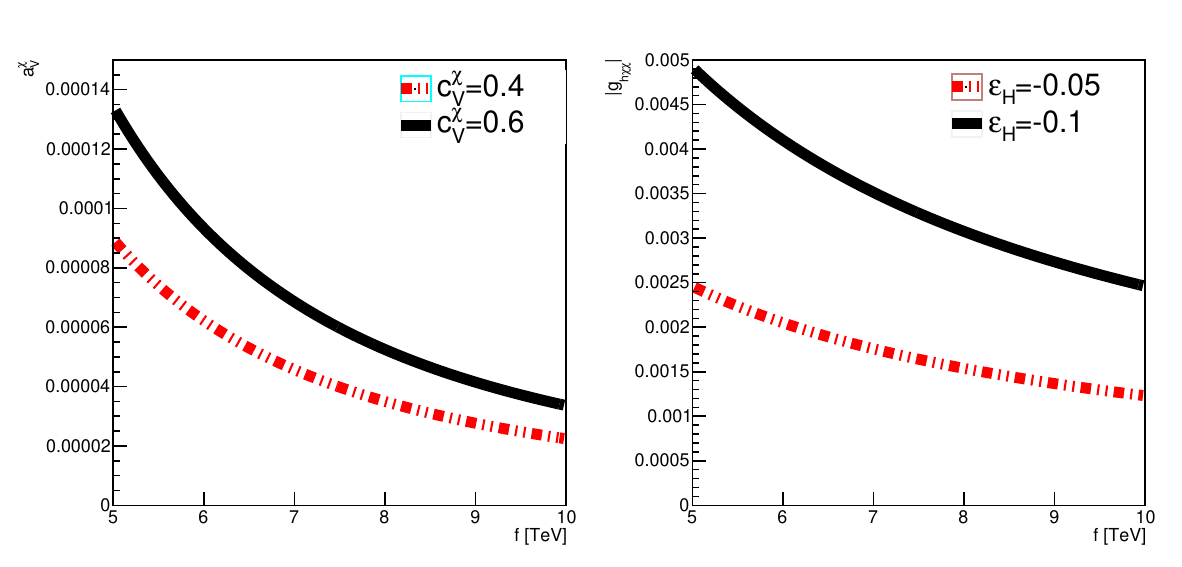}

\caption{The dependence of the couplings $a^{\chi}_{V}$ ({\bf Left}) and $|g_{h\chi\chi}|$ ({\bf Right})
on the compositeness scale $f$. The solid and dotted lines correspond to $c^{\chi}_{V}=0.6$ and $c^{\chi}_{V}=0.4$
({\bf Left}) as well as $\varepsilon_H=-0.1$ and $\varepsilon_H=-0.05$ ({\bf Right}) respectively.}
\label{fig0}
\end{center}
\end{figure}

The coupling of the Higgs state $h$ to nucleons $g_{hNN}$ is set by
\be
g_{hNN} = a^{N}_{S} \dfrac{m_N}{v}\,,
\label{20}
\ee
where
\be
a^{N}_{S} = \sum_{q=u,d,s} f^N_{Tq} + \dfrac{2}{27} \sum_{Q=c,b,t} f^N_{TQ}\,,
\label{22}
\ee
$$
\langle N | m_{q}\bar{q}q |N \rangle= m_N f^N_{Tq}\,, \qquad\qquad\qquad\qquad f^N_{TQ} = 1 - \sum_{q=u,d,s} f^N_{Tq}\,.
$$
The coefficients $f^N_{Tq}$ are related to the $\pi$--nucleon $\sigma$ term and the spin content of the nucleon.
In our analysis we fix $f^p_{Tq}\simeq f^n_{Tq}\simeq f_{Tq}$, i.e. $a^{p}_{S}\simeq a^{n}_{S}\simeq a_{S}$, as well as
$f_{Tu}\simeq 0.0153$, $f_{Td}\simeq 0.0191$ and $f_{Ts}\simeq 0.0447$ which are the default values used in micrOMEGAs~\cite{Belanger:2013oya}.

In the leading approximation the spin--independent $\chi-N$ scattering cross section averaged over initial spins and summed over final spins
takes the form
\begin{equation}
\sigma=\dfrac{m^2_r}{\pi} \Biggl|\dfrac{g_{h\chi\chi} a_{S} m_N}{v m^2_{h}} -
\dfrac{\bar{g} a^{\chi}_{V} \langle a_{V} \rangle }{2 M_Z^2}\Biggr|^2 \,,
\qquad\qquad m_r=\dfrac{m_{\chi} m_N}{m_{\chi}+m_N}\,.
\label{23}
\end{equation}
In Eq.~(\ref{23}) $M_Z$ and $m_h$ are the masses of the $Z$--boson and Higgs scalar, i.e.
$M_Z \simeq 91.2\,\mbox{GeV}$ and $m_h \simeq 125\,\mbox{GeV}$, while
$\langle a_{V} \rangle$ is given by
$$
\langle a_{V} \rangle = \dfrac{1}{A}\Biggl( Z a^{p}_{V} + (A-Z) a^{n}_{V} \Biggr)\,,
$$
where $A$ and $Z$ are the nucleon number and charge of the target nucleus.
Focussing on the case of xenon target nucleus we set $A\approx 130$ and $Z=54$.

\begin{figure}
\begin{center}
	
\hspace*{-2em}\includegraphics[scale=0.9]{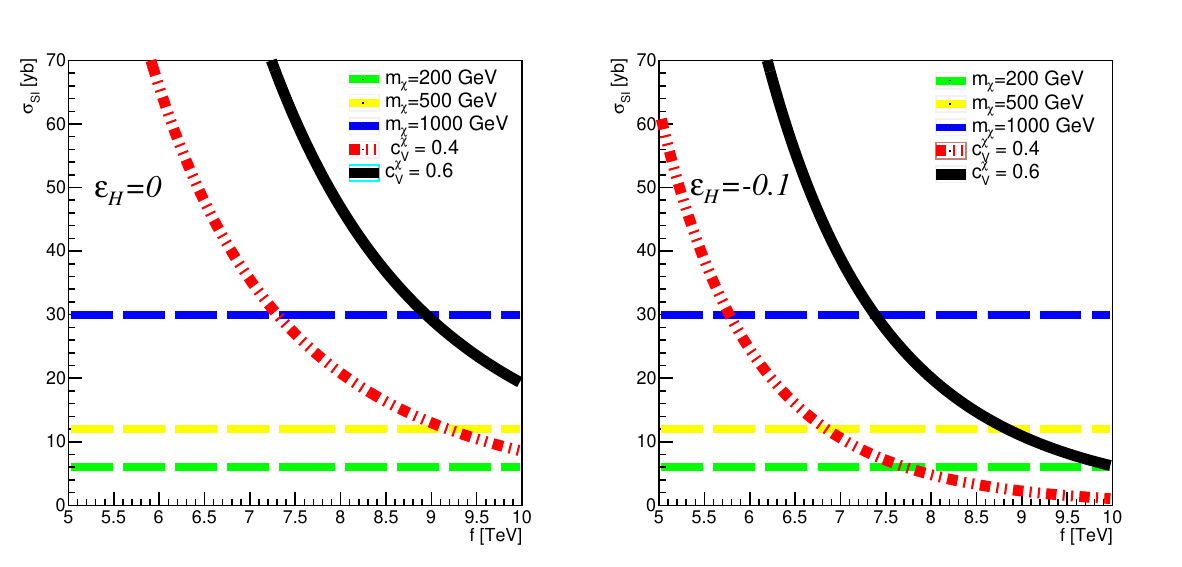}

\caption{The dependence of the spin--independent $\chi-N$ scattering cross section $\sigma$
on the compositeness scale $f$ for $\varepsilon_H=0$ ({\bf Left}) and $\varepsilon_H=-0.1$ ({\bf Right}).
The solid and dotted lines correspond to $c^{\chi}_{V}=0.6$ and $c^{\chi}_{V}=0.4$ respectively.
The highest, second lowest and lowest dashed lines represent the experimental limits on $\sigma$ for
$m_{\chi}=1\,\mbox{TeV}$, $m_{\chi}=500\,\mbox{GeV}$ and $m_{\chi}=200\,\mbox{GeV}$.}
\label{fig1}
\end{center}
\end{figure}

The results of our analysis are summarised in Fig.~2-3. In Fig.~2 the results of
theoretical calculations of the spin--independent $\chi-N$ scattering cross
section in the E$_6$CHM are compared with the corresponding experimental limits
which are $6\,\mbox{yb}$, $12\,\mbox{yb}$ and $30\,\mbox{yb}$ for the dark matter
masses $200\,\mbox{GeV}$, $500\,\mbox{GeV}$ and $1\,\mbox{TeV}$ respectively
\cite{LZ}. Since the E$_6$CHM does not possess the $SU(2)_{cust}$ symmetry we
focus on the part of the parameter space that corresponds to $f\gtrsim 5\,\mbox{TeV}$
as it was discussed in Section 2. To avoid too large fine-tuning associated with
the stabilisation of the EW scale we restrict our consideration to
$f\lesssim 10\,\mbox{TeV}$. Here we also assume that $m_{\chi}\gg m_N$ so that $\sigma$ does
not change much when $m_{\chi}$ varies. As a consequence the spin--independent
$\chi-N$ scattering cross section (\ref{23}) mostly depends on $f$, $c^{\chi}_{V}$ and
$\varepsilon_H$. As follows from Fig.~2 the corresponding cross section diminishes with increasing $f$.
Moreover the spin--independent $\chi$--nucleon scattering cross section also reduces when the
absolute value of $c^{\chi}_{V}$ decreases.
Negative values of $\varepsilon_H$ results in further reduction of $\sigma$.

\begin{figure}
\begin{center}
	
\hspace*{-2em}\includegraphics[scale=0.75]{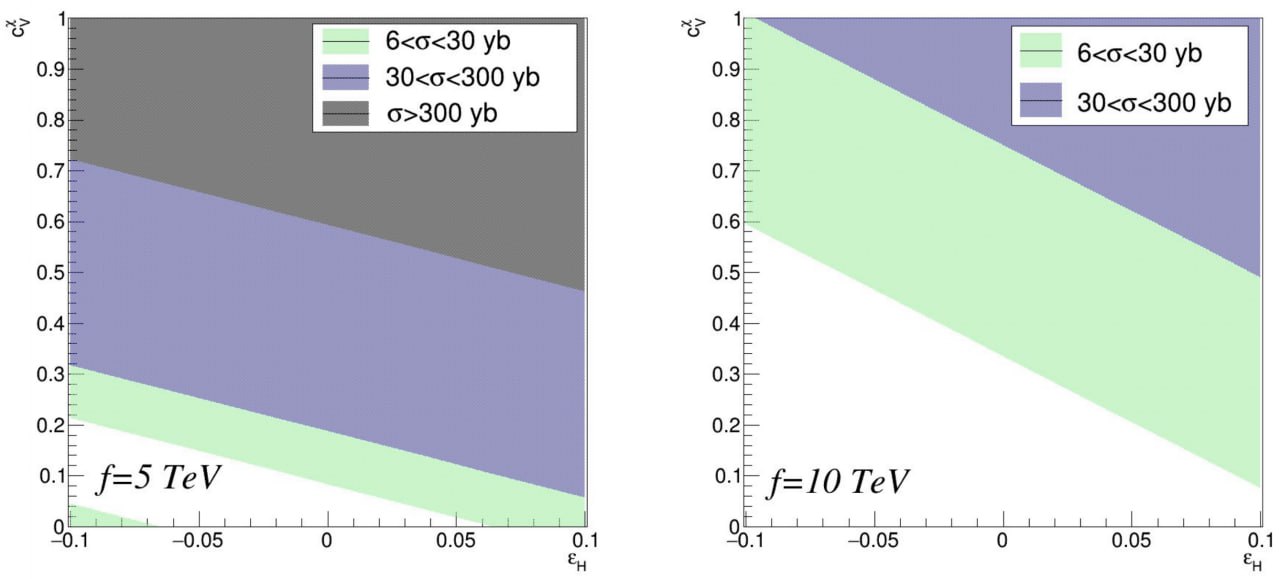}

\caption{Different regions of the E$_6$CHM parameter space in the $\varepsilon_H - c^{\chi}_{V}$ plane
for $f=5\,\mbox{TeV}$ ({\bf Left}) and $f=10\,\mbox{TeV}$ ({\bf Right}). White, green, blue and dark grey
regions correspond to $\sigma < 6\,\mbox{yb}$, $6\,\mbox{yb} < \sigma < 30\,\mbox{yb}$,
$30\,\mbox{yb} < \sigma < 300\,\mbox{yb}$ and $\sigma > 300\,\mbox{yb}$ respectively.}
\label{fig2}
\end{center}
\end{figure}

Because of this for $f\simeq 5\,\mbox{TeV}$ and $\rho_{\chi}\simeq \rho_{DM}$ one can still find the
phenomenologically viable scenarios even if $m_{\chi}\lesssim 200\,\mbox{GeV}$, i.e. $\sigma < 6\,\mbox{yb}$.
This is clearly indicated in Fig.~3. In our analysis we vary $\varepsilon_H$ from $-0.1$ to $0.1$ and
limit ourselves to positive values of $c^{\chi}_{V}$ which are smaller than unity.
The allowed part of the parameter space becomes noticeably
bigger for $m_{\chi}\lesssim 1\,\mbox{TeV}$ ($\sigma < 30\,\mbox{yb}$). For $f\simeq 5\,\mbox{TeV}$ the
constraints on the E$_6$CHM parameter space gets substantially weakened if the lightest exotic fermions
constitute only 10\% of the total dark matter density since in this case $\sigma < 300\,\mbox{yb}$ for
$m_{\chi}\simeq 1\,\mbox{TeV}$ (see Eq.~(\ref{101})). It is much easier to find the phenomenologically
viable scenarios when $f\simeq 10\,\mbox{TeV}$. Fig.~3 demonstrates that for $m_{\chi}\lesssim 200\,\mbox{GeV}$
and $m_{\chi}\lesssim 1\,\mbox{TeV}$ the allowed regions of the E$_6$CHM parameter space become
much wider as compared with $f\simeq 5\,\mbox{TeV}$. In the near future the experiments XENONnT \cite{XENON:2020kmp}
and LUX--ZEPLIN (LZ) \cite{LUX-ZEPLIN:2018poe}, may set even more stringent constraints on the spin--independent
$\chi-N$ scattering cross section.

\section{Conclusions}

Recent measurements set stringent constraints on the magnetic dipole moment $\mu_{\chi}$ of the Dirac DM fermions.
Such fermions may account for all or some of the cold dark matter density
in the composite Higgs models. However the experimental limits on $\mu_{\chi}$ imply that
even if these fermions compose a fraction of DM density larger than $0.1\%$,
the compositeness scale $f$ in the corresponding extensions of the SM
has to be substantially larger than $100\,\mbox{TeV}$ which makes the stabilization
of the EW scale rather problematic. In this article we argued that in the composite Higgs models
the approximate $U(1)$ symmetry can result in the relatively light Dirac DM fermion with suppressed
magnetic dipole moment and small coupling to the $125\,\mbox{GeV}$ Higgs state $h$.

In particular, we considered the $E_6$ inspired composite Higgs model (E$_6$CHM) in which the strongly interacting sector
possesses the approximate $SU(6)$ symmetry and the right--handed top quark is composite. The particle content of the E$_6$CHM
involves a set of exotic fermions which contains two SM singlet states $N_{1}$ and $\overline{N}_{2}$ with baryon numbers
$1/3$ and $(-1/3)$. These states form Dirac fermion $\chi$. The $SU(6)$ symmetry breaking near the scale $f$ down to its
$SU(5)$ subgroup, which contains the SM gauge group, leads to the pNGB states in the E$_6$CHM that form the SM Higgs doublet $H$,
the scalar colour triplet $T$ and the SM singlet $\phi_0$. In this model the color triplet of scalar fields tends to have
mass $m_T$ which is considerably smaller than the compositeness scale $f$, whereas all extra fermions gain masses of the order
of $f$. If the scalar color triplet were lighter than all exotic fermions then it would be stable. Such scenarios are ruled out.
On the other hand, when the E$_6$CHM Lagrangian is invariant under the transformations of global
$U(1)_E$ symmetry, the mass of lightest exotic fermion $m_{\chi}$, its magnetic dipole moment and coupling to $h$ vanish.
The simplest phenomenologically viable scenarios imply that E$_6$CHM possesses the approximate $U(1)_E$ symmetry.
In this case $\chi$ tends to be the lightest exotic fermion in the spectrum and can have mass which is smaller than $m_T$.
The baryon number conservation ensures that the Dirac fermion $\chi$ is stable.

The lower bound on the compositeness scale $f$ in this model caused by the electroweak
precision measurements is about $5\,\mbox{TeV}$. Because of this the coupling of the lightest exotic fermion $\chi$
to $Z$ boson is quite suppressed. Neglecting $\mu_{\chi}$ we explored the spin-independent interactions of $\chi$
with nucleons which are dominated by $t$-channel exchange of the $Z$ boson and the SM Higgs scalar.
It was demonstrated that the present experimental limits set stringent constraints on the E$_6$CHM
parameter space for $f\simeq 5\,\mbox{TeV}$ if $\rho_{\chi}\simeq \rho_{DM}$ and $m_{\chi}\lesssim 1\,\mbox{TeV}$.
Still one can find the phenomenologically viable scenarios even in this case. The constraints become substantially
weaker if $\rho_{\chi}\simeq 0.1\cdot \rho_{DM}$ or the compositeness scale $f$ is considerably larger than $5\,\mbox{TeV}$.

The E$_6$CHM predicts the existence of the scalar colour triplet $T$ that might be lighter than $2\,\mbox{TeV}$
if $f\simeq 5-6\,\mbox{TeV}$. At the LHC such states can be pair--produced resulting in some enhancement
of the cross section of $pp\to b\overline{b} + E^{\rm miss}_{T}$\,.
The discovery of the scalar colour triplet may permit to distinguish the E$_6$CHM from other extensions of the~SM.

\section*{Acknowledgements}

\vspace{-2mm} The authors are grateful to M.~I.~Vysotsky for valuable comments and remarks.


\end{document}